\documentclass[journal]{IEEEtran}

\usepackage{amsmath}
\usepackage{graphicx}
\usepackage{url}
\usepackage{hyperref}
\usepackage{cite}
\usepackage{amsfonts}
\usepackage{mathtools}
\usepackage{hhline}
\usepackage[caption=false, font=footnotesize]{subfig}

\usepackage{algpseudocode}
\usepackage{algorithm}
\algnewcommand\algorithmicforeach{\textbf{for each}}
\algdef{S}[FOR]{ForEach}[1]{\algorithmicforeach\ #1\ \algorithmicdo}

\usepackage[table]{xcolor}
\usepackage{commath,multirow}
\usepackage{tabularx} 
\usepackage{diagbox}

\usepackage{verbatim} 

\usepackage{adjustbox,cleveref}

\newcommand{\etal}{\textit{et al}. }
\newcommand{\ie}{{i}.{e}.}
\newcommand{\eg}{{e}.{g}.}

\makeatletter
\long\def\@makecaption#1#2{\ifx\@captype\@IEEEtablestring%
\footnotesize\begin{center}{\normalfont\footnotesize #1}\\
{\normalfont\footnotesize\scshape #2}\end{center}%
\@IEEEtablecaptionsepspace
\else
\@IEEEfigurecaptionsepspace
\setbox\@tempboxa\hbox{\normalfont\footnotesize {#1.}~~ #2}%
\ifdim \wd\@tempboxa >\hsize%
\setbox\@tempboxa\hbox{\normalfont\footnotesize {#1.}~~ }%
\parbox[t]{\hsize}{\normalfont\footnotesize \noindent\unhbox\@tempboxa#2}%
\else
\hbox to\hsize{\normalfont\footnotesize\hfil\box\@tempboxa\hfil}\fi\fi}
\makeatother

\begin{document}
%
\title{ST-GREED: Space-Time Generalized Entropic Differences for Frame Rate Dependent \\Video Quality Prediction}

\author{Pavan C. Madhusudana, Neil Birkbeck, Yilin Wang,  Balu Adsumilli and Alan C. Bovik 
	\thanks{P. C. Madhusudana and A. C. Bovik are with the Department of Electrical and
Computer Engineering, University of Texas at Austin, Austin, TX, USA (e-mail:
pavancm@utexas.edu; bovik@ece.utexas.edu). Neil Birkbeck, Yilin Wang
and Balu Adsumilli are with Google Inc. (e-mail: birkbeck@google.com; yilin@google.com; badsumilli@google.com).}}


\maketitle

\begin{abstract}
We consider the problem of conducting frame rate dependent video quality assessment (VQA) on videos of diverse frame rates, including high frame rate (HFR) videos. More generally, we study how perceptual quality is affected by frame rate, and how frame rate and compression combine to affect perceived quality. We devise an objective VQA model called Space-Time GeneRalized Entropic Difference (GREED) which analyzes the statistics of spatial and temporal band-pass video coefficients. A generalized Gaussian distribution (GGD) is used to model band-pass responses, while entropy variations between reference and distorted videos under the GGD model are used to capture video quality variations arising from frame rate changes. The entropic differences are calculated across multiple temporal and spatial subbands, and merged using a learned regressor. We show through extensive experiments that GREED achieves state-of-the-art performance on the LIVE-YT-HFR Database when compared with existing VQA models. The features used in GREED are highly generalizable and obtain competitive performance even on standard, non-HFR VQA databases. The implementation of GREED has been made available online: \url{https://github.com/pavancm/GREED}.
\end{abstract}

\begin{IEEEkeywords}
high frame rate, objective algorithm evaluations, video quality assessment, full reference, entropy, natural video statistics, generalized Gaussian distribution
\end{IEEEkeywords}

\section{Introduction}
\IEEEPARstart{P}{roviding} immersive visual experiences for consumers is of principal importance for entertainment, streaming and social video service providers. In recent years considerable effort has been expended on improving video quality by extending current video parameter spaces along spatial and temporal resolutions, color gamut, dynamic range and multiview formats. However there has been less attention directed to high frame rate (HFR) videos, and existing major television, cinema and other video streaming applications currently only deliver videos at 60 frames per second (fps) or less.

The impact of frame rate on perceptual video quality is a less studied topic. Although there exists a belief that HFR videos generally possess superior perceptual quality due to implied reductions in temporal artifacts such as aliasing (flicker, judder etc.) and motion blur, there have been few systematic studies validating these notions. As interest in HFR video delivery has begun to increase, owing to a plethora of high motion and sports and live action content, the question naturally arises whether effective HFR video quality assessment (VQA) databases and measurement tools can be successfully developed and put into practice. 

The problem of HFR-VQA has been previously attempted by analyzing databases like Waterloo HFR \cite{nasiri2015perceptual} and BVI-HFR \cite{mackin2018study}, which primarily address HFR content quality. Although these databases try to address the problem of frame rate dependent video quality prediction, they suffer from some fundamental limitations: only a few frame rates are considered, or the combined effects of compression distortions are not considered. Fortunately, a new HFR-VQA database, called LIVE-YT-HFR \cite{pavan2020liveythfr} has been published, spanning 6 different frame rates ranging from 24 to 120 fps, combined with diverse levels of compression distortions. This publicly available database provides a new and valuable tool to enable the modeling of the complex relationships between frame rate, compression and perceptual video quality.

Existing generic VQA models are of limited use in this application space, since the reference and distorted videos that they compare are required to have the same frame rates. Although existing VQA methods can be extended to HFR scenarios by suitable preprocessing methods, such as temporal downsampling of the reference videos, or upsampling of distorted videos, these operations often result in new distortions and poor correlations against human judgments of video quality \cite{pavan2020liveythfr}. Moreover, the performances of these ``standard'' models can be sensitive to the choice of preprocessing method employed. 

Here we propose a new HFR-VQA model that we call \textbf{G}ene\textbf{R}aliz\textbf{E}d \textbf{E}ntropic \textbf{D}ifference (GREED), which analyzes the statistics of spatial and temporal band-pass filtered video coefficients against statistical and perceptual models of distortion. Important characteristics of the proposed design are as follows:
\begin{enumerate}
    \item The band-pass coefficients are modeled as following a Generalized Gaussian Distribution (GGD). We show that local space-time block entropies can effectively quantify perceptual artifacts that arise from changes in frame rate or compression or both. 
    \item GREED is composed of separate spatial and temporal features. Spatial GREED (SGREED) features are calculated on spatial band-pass coefficients and can only capture spatial distortions, while temporal GREED (TGREED) features are obtained from temporal band-pass responses, which can capture both spatial and temporal impairments. We also show that SGREED and TGREED features account for complementary perceptual quality information.
    \item GREED features are calculated over multiple spatial and temporal subbands, and then combined using a learned regressor to predict quality.  The parameters of the regressor are obtained in a data driven manner, whereby a mapping is learned from GREED features to quality scores using a suitable VQA dataset.
    \item GREED is highly generalizable, and a family of algorithms is designed that vary in the choice of temporal band-pass filter employed. For example choosing a single level Haar filter is equivalent to a simple frame differencing operation, which has been successfully used in various prior VQA models \cite{soundararajan2012video,VMAF2016,bampis2017speed,bampis2018spatiotemporal}.
    \item Models that employ GREED features achieve state-of-the-art performance on the new LIVE-YT-HFR Database. Moreover, these features achieve competitive performance, even on standard VQA databases showing their generalizability to non-HFR scenarios.
\end{enumerate}

The rest of the paper is organized as follows: In Section \ref{sec:prior_work} we discuss prior work on the objective VQA problem. In Section \ref{sec:Objective_VQA} we provide a detailed description of our proposed VQA model. In Section \ref{sec:experiments} we compare and analyze various experimental results comparing GREED against existing VQA models, and we conclude with thoughts for future work in Section \ref{sec:conclusion}.

\section{Related Work}
\label{sec:prior_work}
Objective VQA models are broadly categorized into three groups \cite{chikkerur2011objective}: Full-Reference (FR), Reduced-Reference (RR) and No-Reference (NR). FR VQA models require access to an entire pristine undistorted video along with its degraded version, while RR models operate with limited reference information. NR models predict quality without any knowledge about a reference. This work addresses the problem of quality evaluation when pristine (reference) and distorted sequences may possibly have different frame rates, thus our primarily focus will be on FR and RR VQA methods.

The literature concerning FR-VQA has significantly matured as a multitude of approaches have been proposed over the last decade. One can trivially extend FR Image Quality Assessment (IQA) indices \cite{wang2004image,wang2003multiscale,zhang2011fsim} for application on videos, by predicting the quality of every video frame and using a suitable temporal pooling scheme. Although this procedure can be computationally inexpensive, performance is limited since useful temporal quality information is not effectively employed. The Video Quality Metric (VQM) \cite{pinson2004new} is an early VQA method, which employs losses in the spatial gradients of luminance, along with features based
on the product of luminance contrast and motion. The later VQM-VFD \cite{pinson2014temporal} model is particularly successful at capturing frame delays, and achieves competitive performance on the LIVE-mobile database \cite{moorthy2012video}. The MOVIE index \cite{seshadrinathan2009motion} and a SSIM-based precursor \cite{seshadrinathan2007structural} use the idea of motion tuning by tracking perceptually relevant artifacts along motion trajectories to measure video quality. ST-MAD \cite{vu2011spatiotemporal} index uses spatio-temporal video slices and applies the idea of ``most apparent distortion" \cite{larson2010most} to assess quality. In \cite{you2013attention}, a contrast sensitivity function derived from an attention-driven foveation mechanism is integrated into a wavelet-distortion visibility measure, yielding a full reference video quality model. The proposed models in \cite{ortiz2014full,manasa2016optical} use optical flow characteristics to measure video quality. Spatio-temporal Natural Scene Statistics (NSS) based models such as ST-RRED \cite{soundararajan2012video} and SpEED-VQA \cite{bampis2017speed} compute spatial and temporal entropic differences in a band-pass domain to measure quality deviation. A common principle underlying in these NSS based approaches is that pristine video frames can be well modeled as following Gaussian Scale Mixture (GSM) statistical model and that the presence of distortions results in departures from the GSM model. These statistical deviations can be used to quantify and predict video quality. Recently, data driven approaches have become increasingly popular because of the performances they are able to deliver. For example, Video Multi-method Fusion (VMAF) \cite{VMAF2016} model developed by Netflix is a widely used quality predictor built on existing IQA/VQA models on features which are combined using a Support Vector Regressor (SVR). ST-VMAF \cite{bampis2018spatiotemporal} improves upon VMAF by affixing additional perceptually relevant features that better express temporal aspects of video quality. The recent popularity of deep learning has led to a variety of CNN based models \cite{kim2018deep,becker2019neural,xu2020c3dvqa} that achieve competitive performance on existing VQA databases. Note that all of the above models require the reference and distorted videos to have the same number of frames in temporal synchrony as well as same spatial resolution. Because of this additional non-trivial processing and undesirable processing steps are required when either of these two conditions are violated.  

\begin{figure*}[!t]
    \centering
    \subfloat[Histograms of band-pass video coefficients \label{fig:dist_un_normalize}]{\includegraphics[width=0.5\textwidth]{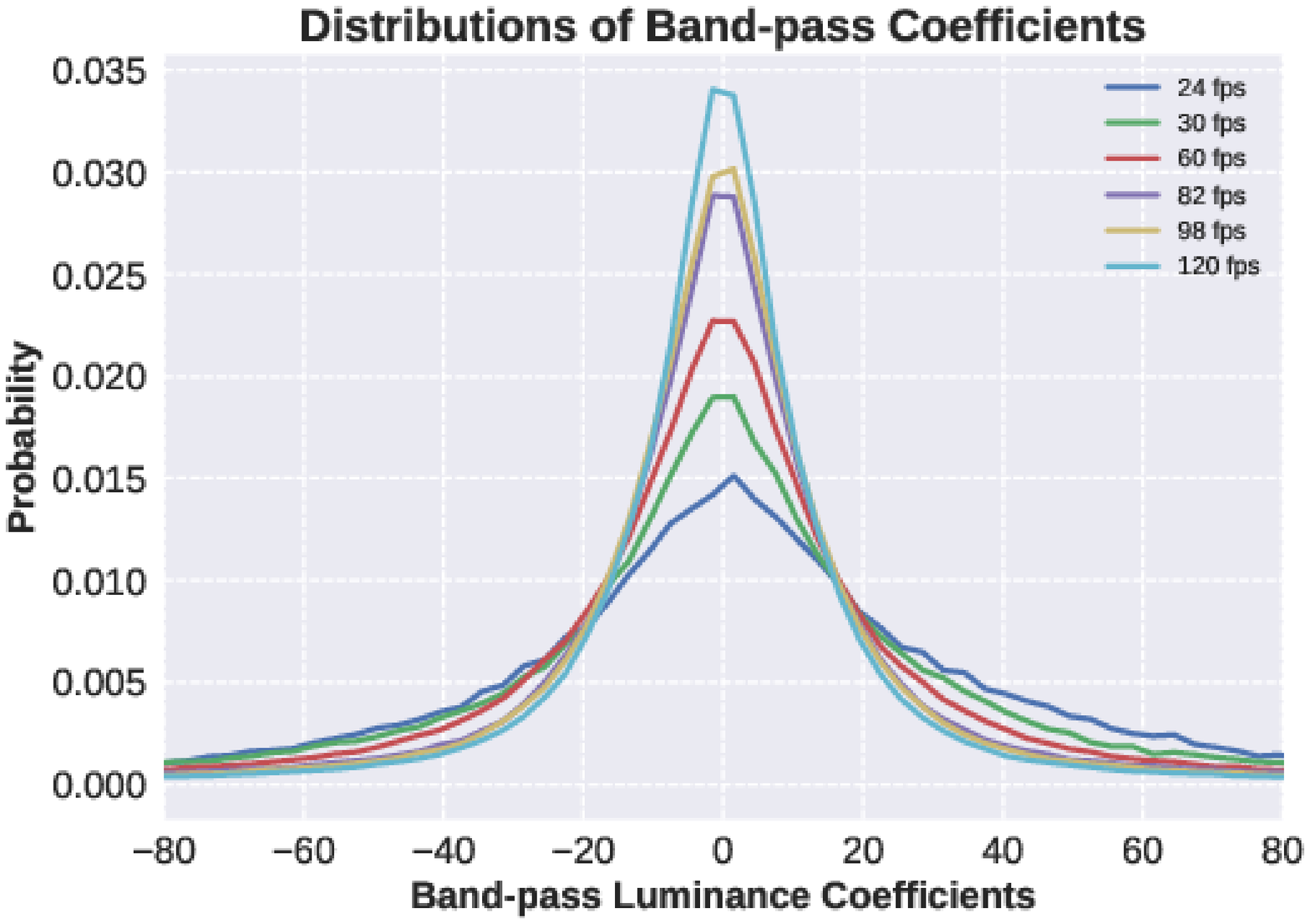}} \hfill
    \subfloat[Histograms of normalized bandpass video coefficients \label{fig:dist_normalized}]{\includegraphics[width=0.5\textwidth]{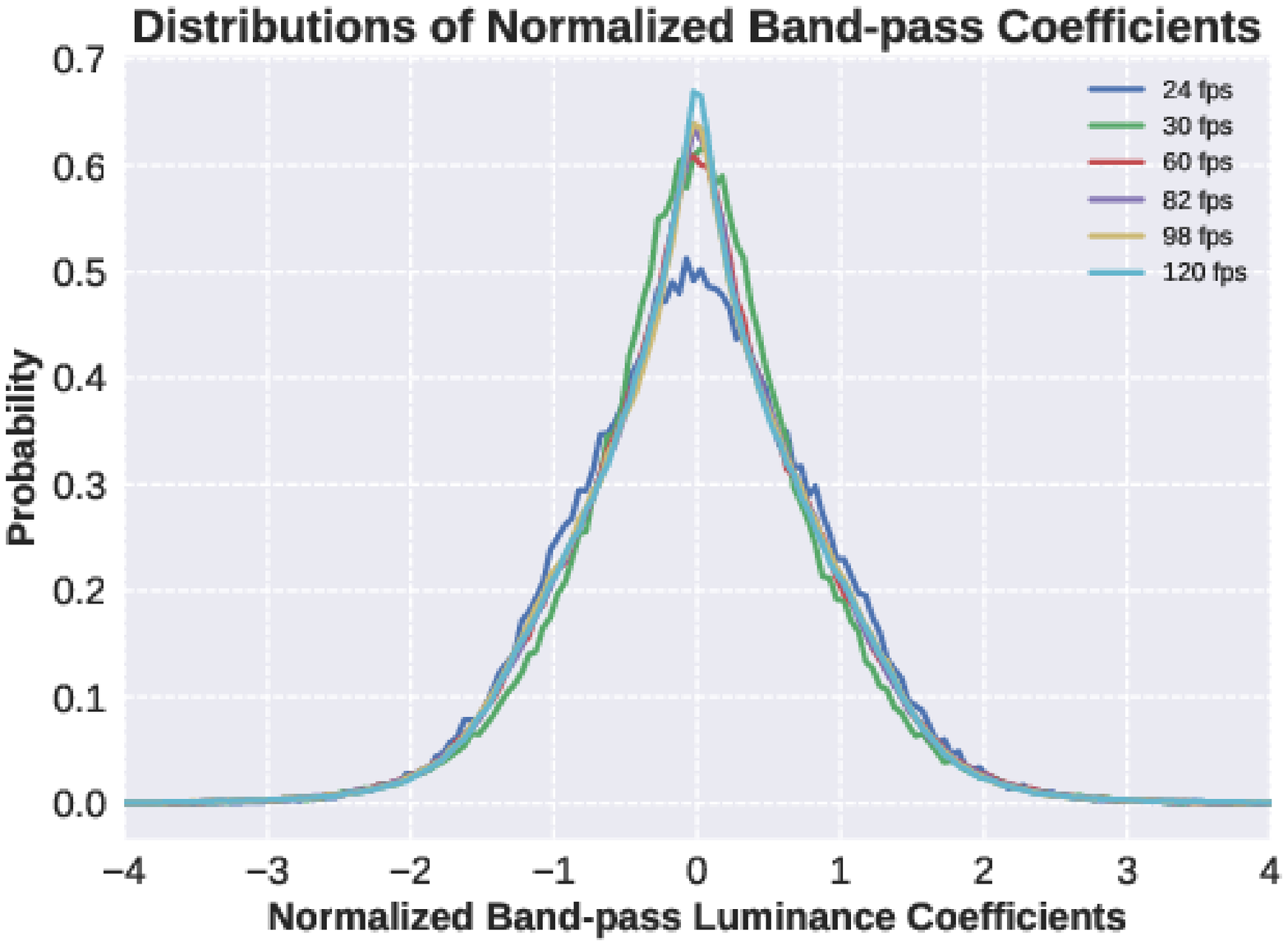}}
    \caption{Comparisons of distributions with and without divisive normalization for \texttt{bobblehead} sequence from LIVE-YT-HFR database.}
    \label{fig:dist_compare}
\end{figure*}

Research pertaining to HFR quality prediction is nascent and the associated literature is very sparse. One of the earliest HFR-VQA models was proposed by Nasiri \etal \cite{nasiri2017perceptual}, where they measure the amount of aliasing in the temporal frequency spectrum to evaluate quality. In \cite{nasiri2018temporal} motion smoothness is measured by examining the local phase correlation of complex wavelet coefficients. This model achieved good performance in the presence of global motion but falls short in presence of local motions or high spatial variations of motion. Zhang \etal \cite{zhang2017frame} proposed the wavelet domain based Frame Rate Quality Metric (FRQM), which uses absolute differences between the temporal wavelet filtered sequences of the reference video and the temporally upsampled distorted video to quantify quality. Although FRQM achieves competitive performance on the BVI-HFR database \cite{mackin2018study}, it cannot be used when the reference and distorted videos have the same frame rate. Moreover, FRQM does not account for the combined effects of compression and frame rate, thus limiting its generalizability.
In previous work \cite{pavan2020gsti}, a temporal Haar filter based GGD model was proposed to measure HFR video quality without including any temporal pre-processing stage, which achieved good performance on the LIVE-YT-HFR Database. Here we substantially extend and generalize this concept by using very general classes of temporal filters, which are deployed over multiple scales, and by using data-driven learning to achieve significantly improved perceptual quality prediction.

\section{Proposed Method}
\label{sec:Objective_VQA}

Here we introduce a novel FR-VQA called GREED, which can be employed when the reference and the distorted videos have either the same or different frame rates. Our model is inspired by prior NSS based models which measure statistical deviations in a bandpass transform domain to quantify quality. These methods rely on two specific principles: One is the strongly decorrelating property of spatial band-pass image decompositions such as the DCT \cite{saad2014blind}, or wavelets \cite{sheikh2006image,soundararajan2012rred,soundararajan2012video} the responses of which strongly tend to follow well-modeled heavy-tailed distributions. The second is subsequent divisive normalization of the bandpass coefficients, the perceptually relevant GSM natural image model and known functional processing by retino-cortical neurons, which relates to contrast masking. Divisive normalization, or equivalently, conditioning on the bandpass variance field \cite{sheikh2006image}, applied on on the bandpass coefficients of images/videos further decorrelates and strongly Gaussianizes them \cite{ruderman1994statistics,mittal2012no}. However, these statistical consistencies tend to be disrupted by distortions which is particularly useful when capturing quality variations. Although there exists a multitude of NSS inspired VQA models, these primarily use spatial NSS models or very simple temporal extensions of them. There has been much less attention directed towards designing temporal NSS models to address temporal artifacts. Current models account for temporal distortions by employing basic operations, such as frame differences \cite{soundararajan2012video,saad2014blind,VMAF2016,bampis2018spatiotemporal} or that perform expensive motion estimation computations \cite{seshadrinathan2009motion,vu2011spatiotemporal,manasa2016optical}. Although these methods perform well under generic conditions, they can only  be used when the reference and distorted videos have the same frame rate, \ie the same number of frames which are in temporal correspondence. Here we aim to go beyond the scope of these prior methods, by removing the frame rate limitation, while addressing the measurement of quality disruptions arising from frame rate variations as well as combined frame rate and compression effects.

Consider a bank of $K$ temporal band-pass filters denoted by $b_k$ for $k \in \{1,\ldots K\}$. The temporal band-pass response to a video $V(\mathbf{x},t)$ ($\mathbf{x} = (x,y)$ represents spatial co-ordinates and $t$ denotes temporal dimension) is denoted by
\begin{align}
    B_k(\mathbf{x},t) = V(\mathbf{x},t)*b_k(t) \text{\hspace{10pt}} \forall k \in \{1,\ldots K\},
    \label{eqn:filter_bank}
\end{align}
where $*$ and $B_k$ are the convolution operation and band-pass response of the $k^{th}$ filter respectively. Note that these are 1D filters applied only along the temporal dimension. We also note that frame differences are a special case of (\ref{eqn:filter_bank}), where the band-pass filter is the high pass component of a Haar wavelet filter. We have empirically observed that the distributions of the coefficients of $B_k$ varies with frame rate. This is illustrated in Fig. \ref{fig:dist_un_normalize} where the empirical distributions (histograms) of bandpass videos having different frame rates are shown following temporal filtering using a 4-level Haar wavelet filter. It may be observed that, as the frame rates increase, the distribution becomes peakier since the correlations between neighborhood frames increase with frame rate, making band-pass responses more sparse. An interesting phenomenon is illustrated in Fig. \ref{fig:dist_normalized}, where the distributions of divisively normalized bandpass coefficients under a GSM model are plotted; any differences between the histograms at are very small. This observation is rather unconventional, given that in many prior models \cite{mittal2012no,mittal2013making,soundararajan2012video,bampis2017speed} it is assumed that divisive normalization tends to capture underlying distortions because they also predict the empirical distributions of the processed image or video signals. One possible explanation behind this observation could be that successful normalization relies on spatial local neighborhood responses, which frame rate changes may not significantly alter. Note that so far, the observations that we are making are in regards to videos having different frame rates, but no compression. Certainly, the presence of compression artifacts can further impact the shape of the distributions.

Although this implies that bandpass normalization may not be strongly predictive of frame rate variations, the band-pass coefficients without normalization are still well modeled as obeying a Generalized Gaussian Distribution (GGD). GGD models have previously been employed to model band-pass coefficients in many applications, such as image denoising \cite{chang2000adaptive}, texture retrieval \cite{do2002wavelet}, blind VQA model \cite{saad2014blind} and so on. Our work is primarily motivated by the successful ST-RRED model \cite{soundararajan2012rred} whereby entropic differences calculated under a GSM model are used to measure deviations in band-pass coefficient distributions caused by distortion. We alter this idea by designing a statistical model based on the GGD, rather than the GSM to capture frame rate variations. In the next subsection, we explain our GGD based model of band-pass coefficients in detail.

\subsection{GGD Based Statistical Model}
\label{subsec:GGD_model}
Let the reference and distorted videos be denoted by $R$ and $D$ respectively, with $R_t,D_t$ representing corresponding frames at time $t$. Note that $R$ and $D$ can possibly have different frame rates, although we will require them to have the same spatial resolution. Let the responses of the $k^{th}$ band pass filter $b_k$, $k\in \{1,2,\ldots K\}$, on the reference and distorted videos be denoted by $B_{kt}^R$ and $B_{kt} ^D$, respectively. Assume that every frame of $B_{kt}^R$, $B_{kt}^D$ follows a GGD model \ie{ } 
\begin{align}
    B_{kt}^R \sim GGD(\mu_{kt} ^{R},\alpha_{kt} ^{R},\beta_{kt} ^{R}), \quad B_{kt}^D \sim GGD(\mu_{kt} ^{D},\alpha_{kt} ^{D},\beta_{kt} ^{D})
    \label{eqn:ggd_assumption}
\end{align}
where $\mu$ is a location parameter which is the mean of the distribution, $\alpha$ is a scale parameter and $\beta$ is the shape parameter. Note that these parameters are time-varying, depending on the dynamics of the video under consideration. Since the band-pass coefficients have zero-mean, we only consider the two parameter GGD model: $\mu_{kt}^{R} = \mu_{kt}^{D} = 0 \text{ }\forall k,t$. The probability density of a zero mean $GGD(\alpha,\beta)$ is given by:
\begin{align*}
    f(x;\alpha,\beta) = \frac{\beta}{2\alpha\Gamma(1/\beta)}\exp\Big(-\Big(\frac{|x|}{\alpha}\Big)^{\beta}\Big)
\end{align*}
where $\Gamma(.)$ is the gamma function:
\begin{align*}
    \Gamma(a) = \int_0 ^{\infty}x^{a-1}e^{-x} dx .
\end{align*}

The shape parameter $\beta$ controls the shape of the distribution (tail weight and peakiness) while $\alpha$ affects the variance. Special cases of GGD include the Gaussian distribution ($\beta = 2$) and Laplacian distribution ($\beta = 1$). Let the band pass coefficients at frame $t$ be partitioned into non-overlapping patches/blocks of size $\sqrt{M} \times \sqrt{M}$, which are indexed by $p \in \{ 1,2, \ldots P \}$. Let $B_{kpt}^R$ and $B_{kpt}^D$ denote vectors of band pass coefficients in patch $p$ for subband $k$ and frame $t$ of the reference and distorted videos, respectively. 

\paragraph*{\textbf{Neural Noise Model}} We model the band-pass coefficients within each patch as having passed through a Gaussian channel, to model perceptual imperfections such as neural noise \cite{sheikh2006image,soundararajan2012video}. Let $B_{kpt}^R,B_{kpt}^D$ represent coefficients which undergo channel imperfections yielding $\Tilde{B}_{kpt}^R,\Tilde{B}_{kpt}^D$. This model is expressed as:
\begin{align}
    \Tilde{B}_{kpt}^R = B_{kpt}^R + W_{kpt}^R, \quad \Tilde{B}_{kpt}^D = B_{kpt}^D + W_{kpt} ^D
    \label{eqn:neural_noise}
\end{align}
where $B_{kpt}^R$ is independent of $W_{kpt}^R$, $B_{kpt}^D$ is independent of $W_{kpt}^D$, $W_{kpt}^R \sim \mathcal{N}(0,\sigma_W ^2 \mathbf{I_M})$ and $W_{kpt}^D \sim \mathcal{N}(0,\sigma_W ^2 \mathbf{I_M})$ and $\mathbf{I_M}$ denotes the identity matrix of dimensions $M \times M$. It may be inferred from (\ref{eqn:neural_noise}) that $\Tilde{B}_{kpt}^R,\Tilde{B}_{kpt}^D$ need not necessarily follow a GGD law, since it is a sum of GGD and Gaussian random variables. However, prior authors \cite{zhao2004sum,soury2015new} have shown that such a sum can be well approximated as GGD under the independence assumption. We hypothesize that the sample entropies of $\Tilde{B}_{kpt}^R$,$\Tilde{B}_{kpt}^D$ contain information potentially pertaining to quality, thus measuring their differences may reflect observed quality differences between the reference and distorted videos. The entropy of a GGD random variable $X \sim GGD(0,\alpha,\beta)$ has a closed form expression given by:
\begin{align}
    h(X) = \frac{1}{\beta} - \log \left(\frac{\beta}{2\alpha\Gamma(1/\beta) }\right).
    \label{eqn:ggd_entropy}
\end{align}

Entropy computation requires knowledge of the GGD parameters of $\Tilde{B}_{kpt}^R$ and $\Tilde{B}_{kpt}^D$. However we only have access to observed band-pass responses, $B_{kpt}^R$ and $B_{kpt}^D$. In order to estimate the parameters of $\Tilde{B}_{kpt}^R$ and $\Tilde{B}_{kpt}^D$ we follow the kurtosis matching procedure detailed in \cite{soury2015new}. The first step involves the estimation of variance, which can be directly calculated from (\ref{eqn:neural_noise}) using the independence assumption:
\begin{equation}
\begin{aligned}
    \sigma^2(\Tilde{B}_{kpt}^R) = \sigma^2(B_{kpt}^R) + \sigma_W ^2, \quad
    \sigma^2(\Tilde{B}_{kpt}^D) = \sigma^2(B_{kpt}^D) + \sigma_W ^2.
\end{aligned}
\label{eqn:ggd_variance}
\end{equation}

The next step involves calculation of the kurtosis $\kappa$:
\begin{equation}
\begin{aligned}
    \kappa(\Tilde{B}_{kpt}^R) &= \kappa(B_{kpt}^R)\left(\frac{\sigma^2(B_{kpt}^R)}{\sigma^2(\Tilde{B}_{kpt}^R)} \right)^2 \\
    \kappa(\Tilde{B}_{kpt}^D) &= \kappa(B_{kpt}^D)\left(\frac{\sigma^2(B_{kpt}^D)}{\sigma^2(\Tilde{B}_{kpt}^D)} \right)^2.
\end{aligned}
\label{eqn:ggd_kurtosis}
\end{equation}

Interested readers can refer to \cite{soury2015new,pan2012exposing} for a detailed derivation of (\ref{eqn:ggd_kurtosis}). The sample variance and kurtosis values of $B_{kpt}^R,B_{kpt}^D$ are employed in (\ref{eqn:ggd_kurtosis}) to calculate the kurtosis of $\Tilde{B}_{kpt}^R$ and $\Tilde{B}_{kpt}^D$, respectively. In the last step, the bijective mapping between the GGD parameters and kurtosis \cite{soury2015new} is applied to estimate the GGD parameters. The expression for the kurtosis of a GGD random variable in terms of its parameters is given by:
\begin{align}
    \kappa(X) = \frac{\Gamma(5/\beta)\Gamma(1/\beta)}{\Gamma(3/\beta)^2}. 
    \label{eqn:param_beta}
\end{align}

A simple grid search can be used to estimate the shape parameter $\beta$ from the kurtosis value obtained from (\ref{eqn:ggd_kurtosis}). The other parameter $\alpha$ is calculated using the relation 
\begin{align}
    \alpha = \sigma \sqrt{\frac{\Gamma(1/\beta)}{\Gamma(3/\beta)}}.
    \label{eqn:param_alpha}
\end{align}

In our implementation we only calculate $\beta$ over each entire frame using (\ref{eqn:param_beta}) rather than on every patch, as we have empirically observed the local predictions to be particularly noisy. However, the scale parameter $\alpha$ is still computed on every patch as it depends on the local variance $\sigma$ along with the shape parameter $\beta$. The entropies $h(\Tilde{B}_{kpt}^R)$ and $h(\Tilde{B}_{kpt}^D)$ are computed by simply substituting the GGD parameters obtained from (\ref{eqn:param_beta}) and (\ref{eqn:param_alpha}), into (\ref{eqn:ggd_entropy}). In the next section we show how these entropies can be effectively used for quality prediction.

\subsection{Temporal Measure}
We define entropy scaling factors similar to the ones used in \cite{soundararajan2012rred,soundararajan2012video} as:
\begin{align*}
    \gamma_{kpt} ^R = \log(1+\sigma^2(\Tilde{B}_{kpt}^R)), \quad \gamma_{kpt}^D = \log(1+\sigma^2(\Tilde{B}_{kpt}^D))
\end{align*}

The notion behind using these scaling factors is to lend more locality to the model. They also have the additional advantage of providing numerical stability on regions having low variance values, as entropy estimates can be noisy/inconsistent in these places. Scaled entropies are obtained by premultiplying these scaling factors:
\begin{align}
    \epsilon_{kpt} ^R = \gamma_{kpt} ^R h(\Tilde{B}_{kpt}^R), \quad \epsilon_{kpt} ^D = \gamma_{kpt} ^D h(\Tilde{B}_{kpt}^D)
    \label{eqn:temporal_scaled_entropy}
\end{align}

\begin{figure}
    \centering
    \includegraphics[width=0.45\textwidth]{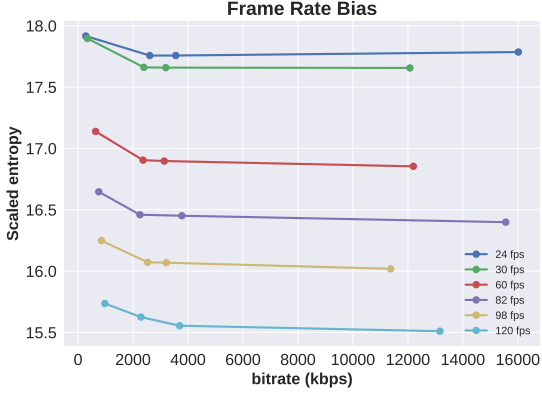}
    \caption{Illustration of frame rate bias of entropy for \texttt{bouncyball} sequence from the LIVE-YT-HFR database. Note that in the above plot, only compressed versions are included, while lossless (CRF = 0) videos are not.}
    \label{fig:frame_rate_bias}
\end{figure}

\paragraph*{\textbf{Frame Rate Bias of Entropy}} Although absolute differences between the scaled entropies in (\ref{eqn:temporal_scaled_entropy}) can represent quality differences, there exists a frame rate bias associated with the entropy values, since different frame rates have entropies at different \textit{scales}. This is illustrated in Fig. \ref{fig:frame_rate_bias}, where the entropy values remain nearly constant at a given frame rate, even with different levels of compression. Thus, simple subtraction can only measure the difference in frame rates between $R$ and $D$. Although this is desirable, it can be ineffective when analyzing videos which only differ by compression artifacts, for example, $R$ and $D$ of the same frame rate but different compression levels. Moreover, quality is jointly dependent on frame rate and compression, while the existence of frame rate bias makes the entropy difference insufficient to capture compression distortions.

To remove this bias, we employ an additional video sequence which we call Pseudo Reference ($PR$) signal, which is obtained by temporally subsampling the reference to match the frame rate of the distorted video. In our experiments, we used the frame dropping technique in FFmpeg \cite{ffmpeg} to achieve this. A similar subsampling scheme was also employed to obtain reduced frame rate videos in the LIVE-YT-HFR database. Note that when a distorted sequence has the same frame rate as the reference, $PR$ is identical to the reference $R$. Similar to $B_{kpt}^R$ and $B_{kpt}^D$, we calculate the band-pass response $B_{kpt}^{PR}$ and its corresponding scaled entropy $\epsilon_{kpt} ^{PR}$. Given these, we define the Temporal-GREED (TGREED) index as:
\begin{align}
    \text{TGREED}_{kt} = \frac{1}{P}\sum_{p=1} ^P \Bigg|\Big(1 + |\epsilon_{kpt} ^D - \epsilon_{kpt} ^{PR}|\Big) \frac{\epsilon_{kpt} ^R + 1}{\epsilon_{kpt} ^{PR} + 1} - 1 \Bigg|
    \label{eqn:GTI}
\end{align}

The expression in (\ref{eqn:GTI}) can be interpreted by breaking down into two components: an absolute difference term and a ratio term. Absolute difference term removes the frame rate bias, while accounting for quality variations as if $R$ and $D$ had the same frame rate. The ratio term acts as a weighting function, where the weights only depend on the frame rates of $R$ and $D$. 

We can derive some very important properties from (\ref{eqn:GTI}). First, when $R$ and $D$ the have same frame rate, the expression depends only on the absolute difference term since the ratio term reduces to 1. The added unit terms within the absolute values ensure that TGREED does not become zero when $D = PR \neq R$, which occurs when the distorted video is a temporally subsampled version of the reference. Note that TGREED = 0 only when $D = PR = R$. The added unit terms in the ratio ensure that indeterminate values of the ratio will not occur in regions having small entropy values.

\subsection{Spatial Measure}
TGREED primarily addresses temporal artifacts by analyzing the statistics of temporal band-pass responses. Although TGREED is calculated in a spatial block based manner, it is mainly influenced by temporal filtering. In order to measure artifacts that arise only or primarily from spatial inconsistencies, we employ spatial band-pass filters applied on every frame of the reference and distorted sequences. To obtain the spatial band-pass responses, we use a simple local Mean Subtracted (MS) filtering scheme similar to \cite{bampis2017speed}. Let $R_t ^{MS} = R_t - \mu_t ^R$ and $D_t ^{MS} = D_t - \mu_t ^D$ be the reference and distorted MS coefficients at frame $t$, where the local mean is calculated as
\begin{align}
\begin{aligned}
    \mu_t ^R(i,j) = \sum_{g=-G} ^G \sum_{h=-H} ^H \omega_{g,h} R_t(i+g,j+h), \\ \mu_t ^D(i,j) = \sum_{g=-G} ^G \sum_{h=-H} ^H \omega_{g,h} D_t(i+g,j+h),
\end{aligned}
\end{align}
where $\omega = \{\omega_{g,h}|g = -G,\ldots G, h = -H,\ldots H\}$ is a 2D circularly symmetric Gaussian weighting function sampled out to 3 standard deviations and rescaled to unit volume, and where $R_t$ and $D_t$ are single frames at time $t$. In our implementation, we use $G = H = 7$. Similar to the temporal case, the spatial band-pass coefficients $R_t ^{MS}$ and $D_t ^{MS}$ are modeled as following GGD law. Again, we divide each frame into non-overlapping patches of size $\sqrt{M} \times \sqrt{M}$, indexed by $p \in \{ 1,2, \ldots P \}$. The channel imperfections can be similarly modeled as:
\begin{align}
    \Tilde{R}_{pt} ^{MS} = R_{pt}^{MS} + Z_{pt}^R, \quad \Tilde{D}_{pt}^{MS} = D_{pt}^{MS} + Z_{pt}^D
    \label{eqn:neural_noise_spatial}
\end{align}
where $R_{pt}^{MS}$ is independent of $Z_{pt}^R$ and $R_{pt}^{MS}$ is independent of $Z_{pt}^D$ and where $Z_{pt}^R \sim \mathcal{N}(0,\sigma_Z ^2 \mathbf{I_M})$ and $Z_{pt}^D \sim \mathcal{N}(0,\sigma_Z ^2 \mathbf{I_M})$. The spatial entropies $h(\Tilde{R}_t ^{MS})$ and $h(\Tilde{D}_t ^{MS})$ are calculated using the procedure detailed in subsection \ref{subsec:GGD_model}, but replacing temporal band-pass responses with MS coefficients. Similarly, we define the scaling factors and modified entropies:
\begin{align}
\begin{aligned}
    \eta_{pt} ^R = \log(1+\sigma^2(\Tilde{R}_{pt}^{MS}))&, \quad \eta_{pt}^D = \log(1+\sigma^2(\Tilde{D}_{pt}^{MS})) \\
    \theta_{pt} ^R = \eta_{pt} ^R h(\Tilde{R}_{pt}^{MS})&, \quad \theta_{pt} ^D = \eta_{pt} ^D h(\Tilde{D}_{pt}^{MS}).
    \label{eqn:spatial_entropy_scaled}
    \end{aligned}
\end{align}
Since spatial entropies are computed using only information from single frames, the obtained values are frame rate agnostic. Thus there does not arise any scale bias due to frame rate as was observed in the temporal case. We define the Spatial-GREED (SGREED) index as:
\begin{align}
    \text{SGREED}_t = \frac{1}{P}\sum_{p=1} ^P |\theta_{pt} ^D - \theta_{pt} ^R|.
    \label{eqn:GSI}
\end{align}

\begin{algorithm}[t] \caption{Generalized Entropic Difference (GREED)}
\begin{algorithmic}[1]
\Require reference video $R$, distorted video $D$
\Ensure GREED score
\State Scale $S = \{4,5\}$, band-pass filter bank $b_k: k \in \{1,\ldots 7\}$
\State Temporal subsample $R$ to get $PR$
\ForEach {$s \in S $}
\State Calculate SGREED from (\ref{eqn:GSI})
\ForEach {$b_k$}
\State Calculate TGREED$_k$ from (\ref{eqn:GTI})
\EndFor
\EndFor
\State Concatenate SGREED, TGREED from all scales in (\ref{eqn:GREED}) to obtain GREED.
\end{algorithmic}
\label{alg1}
\end{algorithm}

\subsection{Spatio-Temporal Measure}
The expressions in (\ref{eqn:GTI}) and (\ref{eqn:GSI}) calculate entropic differences at the frame level. We combine these frame level differences by average pooling over all temporal coordinates, to obtain video level differences:
\begin{align*}
    \text{TGREED}_k &= \frac{1}{T} \sum_{t = 1} ^T \text{TGREED}_{kt}, \\
    \text{SGREED} &= \frac{1}{T} \sum_{t = 1} ^T \text{SGREED}_{t}.
\end{align*}
The factors SGREED and TGREED operate individually on data obtained by separately processing spatial and temporal band-pass responses. Interestingly, while SGREED is obtained in a purely spatial manner, TGREED has both spatial and temporal information embedded in it (since the entropies are obtained in a spatial blockwise manner). Thus temporal artifacts such as judder/strobing only influence TGREED, while spatial artifacts affect both TGREED and SGREED. The combined spatio-temporal GREED index is obtained as a function of both SGREED and TGREED:
\begin{align}
    \text{GREED} = f(\text{SGREED},\text{TGREED}),
    \label{eqn:GREED}
\end{align}
where $f$ is a function which takes SGREED and TGREED as input features and predict quality scores. The mapping $f$ may be learned from a suitable VQA database containing human quality judgments. 

\subsection{Regression}
We employed a Support Vector Regressor (SVR) \cite{scholkopf2000new} that was trained on the LIVE-YT-HFR database, using Difference Mean Opinion Scores (DMOS) to obtain the mapping function $f$ described in (\ref{eqn:GREED}). The SVR is a well known modeling technique which is widely used in many prior IQA/VQA methods \cite{VMAF2016, bampis2018spatiotemporal, mittal2012no, saad2014blind} and is known for obtaining effective non-linear mappings on high-dimensional data with high accuracy. During training, SGREED and TGREED features were calculated on each video and then fed as input to SVR along with the corresponding DMOS labels. In our implementation, we used the LIBSVM \cite{chang2011libsvm} package with radial basis function (RBF) kernel to train and test GREED.

\subsection{Implementation Details}
For simplicity, we implemented our method only in the luminance domain. To accomplish the temporal band-pass filtering, we experiment with 3 wavelet filters: Haar, Daubechies-2 (db2) and Biorthogonal-2.2 (bior2.2). We used wavelet packet (constant linear bandwidth) filter bank \cite{coifman1992entropy} as we found it to be more effective than using constant octave bandwidth filters. The choice of linear bandwidth was also beneficial when analyzing the impacts of individual frequency bands on perceived quality. We employed 3 levels of wavelet decomposition for all wavelet filters $b_k$, $k \in \{1,\ldots 7\}$ (ignoring the low pass response), where higher values of $k$ correspond to filter with higher center frequencies. When calculating entropies we used spatial patches of size $5 \times 5$ (\ie{ }$\sqrt{M} = 5$). The neural noise variance was fixed at $\sigma_W ^2 = \sigma_Z ^2 = 0.1$ in (\ref{eqn:neural_noise}) and (\ref{eqn:neural_noise_spatial}), matching those employed in \cite{soundararajan2012video} and \cite{bampis2017speed}. 

In our experiments we found that our algorithm is most effective when the SGREED and TGREED features are calculated over multiple spatial resolutions. In particular, scales $s = 4$ and $s = 5$ were observed to provide superior performance, where the spatial resolution was downsampled $2^s$ times along both dimensions. The importance of using features from lower scales is likely attributable to the motion downshifting phenomenon, which posits that in the presence of motion, humans tend to be more sensitive to coarser scales than finer ones. Note that similar observations were made in \cite{soundararajan2012video,bampis2017speed,bampis2018spatiotemporal}. Downsampling delivers the additional advantage of significantly reducing the computational complexity. Since each scale results in an 8-dimensional feature vector (one value of SGREED, and TGREED values from each of seven subbands), employing two scales leads to a 16-dimensional vector as the input in (\ref{eqn:GREED}). 

Since reference and distorted sequences can have different frame rates, the reference entropy terms $\epsilon_{kpt} ^R$, $\theta_{pt} ^R$ will generally have a different number of frames when compared to their counterpart distorted entropy terms $\epsilon_{kpt} ^D$, $\theta_{pt} ^D$. Thus we average reference entropy terms over all temporal indices:


\begin{equation*}
\begin{aligned}
{\epsilon}_{kpt}^R &\leftarrow \frac{1}{F}\sum_{n=1}^F \epsilon_{kpt'}^R
 &\quad &\raisebox{-1.5\normalbaselineskip}[0pt][0pt]{
     where \text{$
     \begin{cases}
        F &= \frac{FPS_{\textrm{ref}}}{FPS_{\textrm{dist}}}, \\
        t' &= (t - 1)F + n
      \end{cases}$
    }} \\
{\theta}_{pt}^R &\leftarrow \frac{1}{F}\sum_{n=1}^F \theta_{pt'}^R
\end{aligned}
\end{equation*}

The above procedure is equivalent to dividing the entropy terms into subsequences of length $F$ along the temporal dimension, and averaging each subsequence \cite{mackin2018study}. This results in an equal number of entropy terms from the reference and distorted videos, which can then be used to calculate SGREED and TGREED in (\ref{eqn:GSI}) and (\ref{eqn:GTI}). The entire GREED flow is summarized in Algorithm \ref{alg1}.

\begin{table}[t]
\caption{Performance comparison of GREED against different FR algorithms on the LIVE-YT-HFR Database. In each column, the first and second best models are boldfaced.}
    \label{Table:MOS_comparison}
    \centering
    \begin{tabular}{|c||c|c|c|c|}
        \hline
        ~    & SROCC $\uparrow$ & KROCC $\uparrow$ & PLCC $\uparrow$ & RMSE $\downarrow$ \\ \hline \hline
        PSNR & 0.7802 & 0.5934 & 0.7481 & 7.75 \\ 
        SSIM \cite{wang2004image} & 0.5566 & 0.4042 & 0.5418 & 9.99 \\ 
        MS-SSIM \cite{wang2003multiscale} & 0.5742 & 0.4135 & 0.5512 & 10.01 \\ 
        FSIM \cite{zhang2011fsim} & 0.6528 & 0.4881 & 0.6332 & 9.34 \\ 
        ST-RRED \cite{soundararajan2012video} & 0.6394 & 0.4516 & 0.6073 & 9.58 \\ 
        SpEED \cite{bampis2017speed} & 0.6051 & 0.4437 & 0.5206 & 10.28 \\ 
        FRQM \cite{zhang2017frame} & 0.5133 & 0.3701 & 0.5017 & 10.38 \\ 
        VMAF \cite{VMAF2016}& 0.7782 & 0.5918 & 0.7419 & 8.10 \\
        deepVQA \cite{kim2018deep} & 0.4331 & 0.3082 & 0.3996 & 10.87 \\ \hline
        GREED-Haar & 0.8305 & 0.6389 & 0.8467 & 6.22 \\
        GREED-db2 & \textbf{0.8347} & \textbf{0.6447} & \textbf{0.8478} & \textbf{6.21} \\
        GREED-bior2.2 & \textbf{0.8822} & \textbf{0.7046} & \textbf{0.8869} & \textbf{5.48} \\
        \hline
    \end{tabular}
\end{table}

\section{Experiments and Results}
\label{sec:experiments}
We conduct a series of experiments to evaluate the performance of GREED. We will first describe the experimental settings, comparison methods and basic evaluation criteria. Then we explain how we evaluated GREED against existing state-of-the-art VQA models on the LIVE-YT-HFR database. We conduct a variety of ablation studies to analyze the significance and generalizabiliy of each conceptual feature present in GREED. Additionally, we demonstrate the generalizability of the GREED features by testing them on generic VQA databases. We also report the time complexity associated with GREED.

\begin{table*}[t]
	\caption{Performance comparison of GREED against various FR methods for individual frame rates on the LIVE-YT-HFR Database.}
	\label{Table:FPS_comparison}
	\centering
	\scriptsize
	\scalebox{0.88}{
		\begin{tabular}{|c||c|c|c|c|c|c|c|c|c|c|c|c|c|c|}
			\hline
			& \multicolumn{2}{|c|}{24 fps} & \multicolumn{2}{|c|}{30 fps} & \multicolumn{2}{|c|}{60 fps} & \multicolumn{2}{|c|}{82 fps} & \multicolumn{2}{|c|}{98 fps} & \multicolumn{2}{|c|}{120 fps} & \multicolumn{2}{|c|}{Overall} \\
			\cline{2-15}
			~ & SROCC$\uparrow$ & PLCC$\uparrow$ & SROCC$\uparrow$ & PLCC$\uparrow$ & SROCC$\uparrow$ & PLCC$\uparrow$ & SROCC$\uparrow$ & PLCC$\uparrow$ & SROCC$\uparrow$ & PLCC$\uparrow$ & SROCC$\uparrow$ & PLCC$\uparrow$ & SROCC$\uparrow$ & PLCC$\uparrow$\\ \hline \hline
			PSNR & 0.5411 & 0.4873 & 0.5643 & 0.5352 & \textbf{0.7536} & 0.6945 & \textbf{0.7714} & 0.7714 & 0.8214 & 0.7834 & 0.7413 & 0.7364 & 0.7802 & 0.7481 \\ 
			SSIM \cite{wang2004image} & 0.2661 & 0.2228 & 0.2839 & 0.1892 & 0.3821 & 0.3027 & 0.3714 & 0.3621 & 0.5375 & 0.4975 & \textbf{0.8671} & \textbf{0.8337} & 0.5566 & 0.5418 \\ 
			MS-SSIM \cite{wang2003multiscale} & 0.3054 & 0.2608 & 0.2964 & 0.2383 & 0.4161 & 0.3382 & 0.4393 & 0.3938 & 0.5786 & 0.5619 & 0.7063 & 0.6963 & 0.5742 & 0.5512\\ 
			FSIM \cite{zhang2011fsim} & 0.3107 & 0.3148 & 0.3161 & 0.3279 & 0.5857 & 0.5083 & 0.4571 & 0.4796 & 0.6839 & 0.6774 & 0.7622 & 0.7062 & 0.6528 & 0.6332\\ 
			ST-RRED \cite{soundararajan2012video} & 0.3054 & 0.2756 & 0.2964 & 0.2066 & 0.6125 & 0.6136 & 0.5839 & 0.5130 & 0.6500 & 0.6042 & 0.7552 & 0.6966 & 0.6394 & 0.6073\\ 
			SpEED \cite{bampis2017speed} & 0.4321 & 0.2729 & 0.4107 & 0.2332 & 0.4393 & 0.2927 & 0.5464 & 0.3901 & 0.5786 & 0.4713 & 0.7587 & 0.7393 & 0.6051 & 0.5206\\ 
			FRQM \cite{zhang2017frame} & 0.2893 & 0.1536 & 0.3196 & 0.1851 & 0.2109 & 0.0999 & 0.2857 & 0.0780 & 0.2750 & 0.0953 & - & - & 0.5133 & 0.5017\\ 
			VMAF \cite{VMAF2016} & 0.2500 & 0.3685 & 0.3625 & 0.4716 & 0.6304 & 0.6806 & 0.7339 & 0.7928 & \textbf{0.8607} & \textbf{0.8685} & 0.8182 & 0.8166 & 0.7782 & 0.7419\\ 
			deepVQA \cite{kim2018deep} & 0.2732 & 0.1576 & 0.2893 & 0.1428 & 0.4036 & 0.2071 & 0.2929 & 0.2401 & 0.4107 & 0.3753 & 0.7622 & 0.6702 & 0.4331 & 0.3996\\ \hline
			GREED-Haar & 0.6196 & 0.6917 & 0.5482 & 0.7646 & 0.7125 & 0.8092 & 0.7464 & 0.8546 & 0.8054 & 0.8491 & 0.8112 & 0.8235 & 0.8305 & 0.8467\\
			GREED-db2 & \textbf{0.6696} & \textbf{0.7396} & \textbf{0.6179} & \textbf{0.7983} & 0.6982 & \textbf{0.8244} & 0.7250 & \textbf{0.8671} & 0.7518 & 0.8449 & 0.8322 & 0.8632 & \textbf{0.8347} & \textbf{0.8478}\\
			GREED-bior2.2 & \textbf{0.7268} & \textbf{0.8221} & \textbf{0.7018} & \textbf{0.8431} & \textbf{0.7321} & \textbf{0.8405} & \textbf{0.8179} & \textbf{0.8960} & \textbf{0.8643} & \textbf{0.8915} & \textbf{0.8881} & \textbf{0.8952} & \textbf{0.8822} & \textbf{0.8869}\\
			\hline
		\end{tabular}
	}
\end{table*}

\subsection{Experimental Settings}
\textbf{Compared Methods}.
Since our proposed framework is an FR/RR model, we selected 4 popular and widely used FR-IQA methods: PSNR, SSIM \cite{wang2004image}, MS-SSIM \cite{wang2003multiscale} and FSIM \cite{zhang2011fsim} for comparison. Since these are IQA models, they do not take into account any temporal information. They were computed on every frame and averaged across all frames to obtain final video scores. In addition to the above IQA metrics, we also included 5 popular FR-VQA models: ST-RRED \cite{soundararajan2012video}, SpEED \cite{bampis2017speed}, FRQM\cite{zhang2017frame}, VMAF\footnote{We used the pretrained VMAF model available at \url{https://github.com/Netflix/vmaf}} \cite{VMAF2016} and deepVQA \cite{kim2018deep}. When evaluating deepVQA, we only used stage-1 of the pretrained model (trained on the LIVE-VQA \cite{seshadrinathan2010study} database) obtained from the code released by the authors. All of the above methods other than FRQM require the reference and corresponding distorted sequences to have the same frame rate. When there were differences in frame rates, we performed naive temporal upsampling by frame duplication to match the reference frame rate. Another way of matching the frame rates is to downsample the reference video, however we did not use this method since it often introduces temporal artifacts in the reference video, such as stutter/judder which is undesirable. We also ignored highly specific temporal upsampling methods (\eg{} motion compensated temporal interpolation), as the performances of these can be highly susceptible to the type of content considered and to the choice of interpolation method.

\textbf{Evaluation Criteria}.
We employed Spearman's rank order correlation coefficient (SROCC), Kendall's rank order correlation coefficient (KROCC), Pearson's linear correlation coefficient (PLCC), and root mean squared error (RMSE) to evaluate the VQA models. All of the above metrics are calculated using DMOS values of the VQA dataset as ground truth. Before computing PLCC and RMSE, the predicted scores were passed through a four-parameter logistic non-linearity as described in \cite{VQEG2000}:
\begin{align}
    Q(x) = \beta_2 + \frac{\beta_1 - \beta_2}{1 + \exp\Bigg(-\Big(\frac{x - \beta_3}{|\beta_4|}\Big)\Bigg)}.
    \label{eqn:logistic_non}
\end{align}

\begin{figure}[t]
    \centering
    \includegraphics[width=0.48\textwidth]{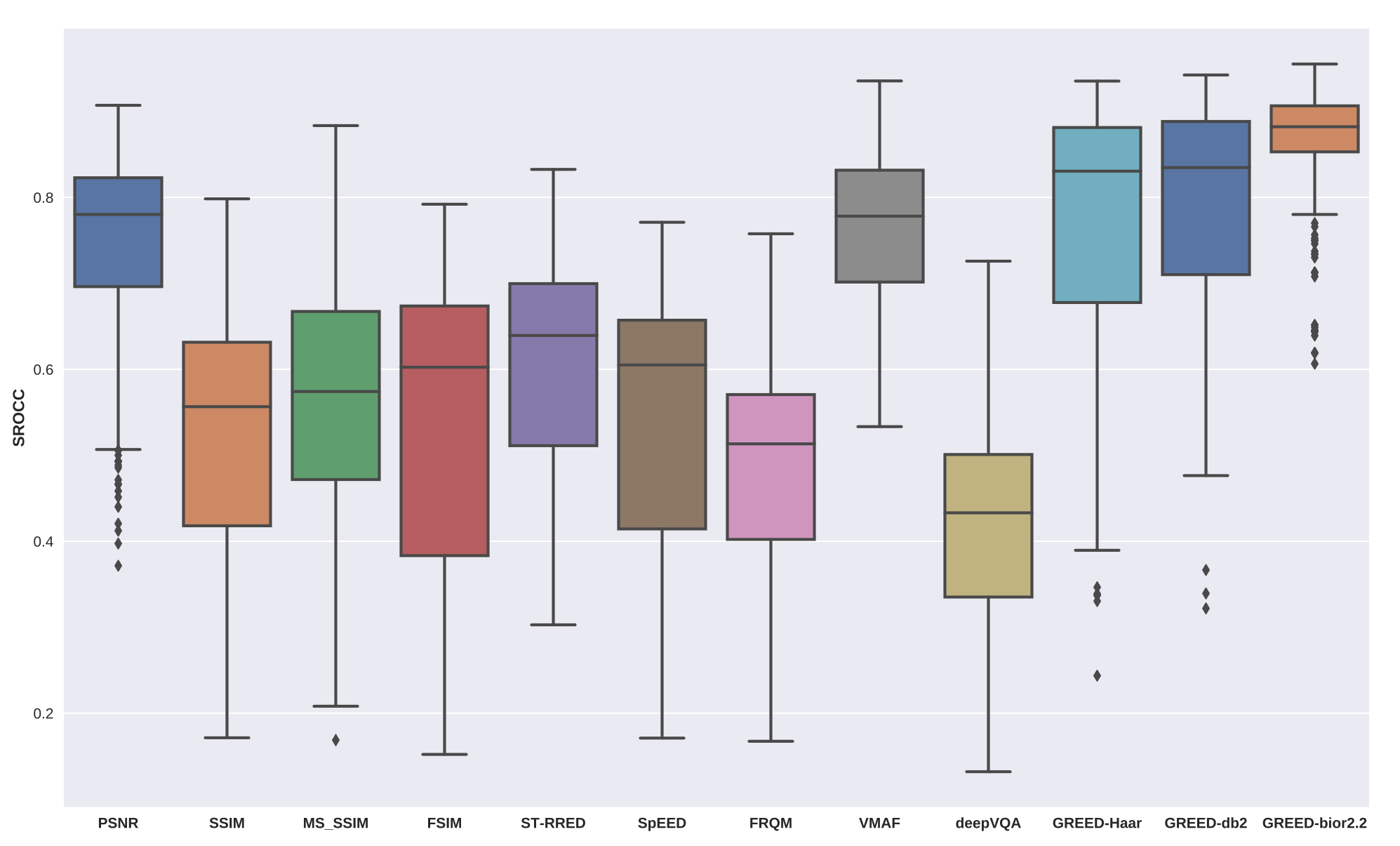}
    \caption{Boxplot of SROCC distributions of the compared VQA algorithms in Table \ref{Table:MOS_comparison} over 200 trials on the LIVE-YT-HFR Database.}
    \label{fig:FR_boxplot}
\end{figure}

\subsection{Correlation Against Human Judgments}
Since GREED requires training on the quality prediction problem, we randomly divide the LIVE-YT-HFR dataset into 70\%,15\%, and 15\% subsets corresponding to training, validation and test sets, respectively. We observed that this choice of train-validation-test splits yielded stable model learning without overfitting/underfitting the training data. We also ensured that there was never any overlap between contents in each set. Since LIVE-YT-HFR comprises 16 contents with 30 videos per content, this implies splits of about 300/90/90 videos in each set. The validation set was used to determine the hyperparameters of the SVR using grid search. We repeated this random train-test sequence over 200 times, and report the median performance.

We compared the performance of GREED against other FR models in Table \ref{Table:MOS_comparison}. It may be observed from the Table that the family of GREED based models significantly outperformed the compared VQA models by a large margin, with GREED-bior2.2 achieving top performance. Fig. \ref{fig:FR_boxplot} shows the spreads of SROCC values for each FR model over 200 iterations. The plot indicates that GREED-bior2.2 has a much tighter confidence interval than the other indices, highlighting the robustness of the algorithm.

In order to individually analyze the performance of GREED against each frame rate we divided the LIVE-YT-HFR database into sets containing videos having the same frame rates. The SROCC and PLCC performance comparison is shown in Table \ref{Table:FPS_comparison}. The KROCC and RMSE were observed to follow the similar trends as in Table \ref{Table:FPS_comparison}. Here as well, the GREED-bior2.2 was among the top two performing models at every frame rate. Note that FRQM requires compared videos to have different frame rates, thus for 120 fps videos correlation values are not reported in Table \ref{Table:FPS_comparison}. We also observed an interesting trend among GREED models whereby the correlation values roughly follow a monotonically increasing behavior with increasing frame rate values. This behavior is expected and desirable, since higher frame rate videos offer more information about the distorted video, thus resulting in better quality judgments.

\begin{table}[t]
\caption{Performance of Spatial and Temporal Measures when evaluated in isolation on the LIVE-YT-HFR Database.}
    \label{Table:individual_component}
    \centering
    \footnotesize
    \begin{tabular}{|c||c|c|c|c|}
        \hline
        ~    & SROCC $\uparrow$ & KROCC $\uparrow$ & PLCC $\uparrow$ & RMSE $\downarrow$ \\ \hline \hline
        SGREED & 0.7233 & 0.5269 & 0.6888 & 8.51 \\ 
        TGREED-Haar & 0.6637 & 0.4881 & 0.6907 & 8.60 \\ 
        TGREED-db2 & 0.5543 & 0.3988 & 0.6121 & 9.49 \\
        TGREED-bior2.2 & 0.5849 & 0.4165 & 0.6180 & 9.41 \\ \hline
        GREED-Haar & 0.8305 & 0.6389 & 0.8467 & 6.22 \\
        GREED-db2 & 0.8347 & 0.6447 & 0.8478 & 6.21 \\
        GREED-bior2.2 & 0.8822 & 0.7046 & 0.8869 & 5.48 \\
        \hline
    \end{tabular}
\end{table}

\subsection{Significance of Spatial and Temporal Measures}
We conduct an ablation study to evaluate the importance of SGREED and TGREED when employed in isolation. In this experiment we separately trained SGREED and TGREED, and the performance values reported in Table \ref{Table:individual_component}. We can infer from the Table that SGREED and TGREED capture complementary perceptual information, since SGREED and TGREED yield lower correlation when used separately, while the combined GREED model obtained much higher performance.

\begin{figure}[t]
    \centering
    \includegraphics[width = 0.48\textwidth]{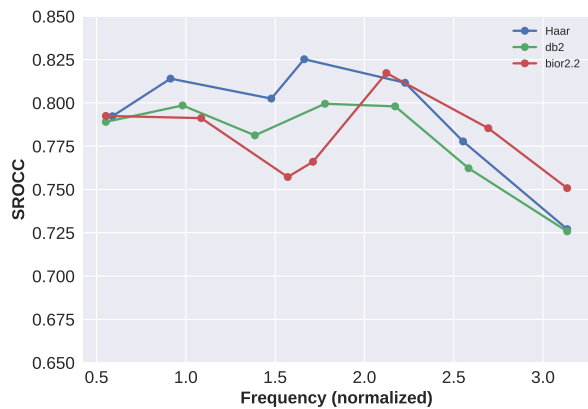}
    \caption{Contribution analysis of each subband present in GREED. Horizontal axis represents center frequencies of each subband normalized to $[0,\pi]$ range.}
    \label{fig:subband_overall}
\end{figure}

\subsection{Contribution of Temporal Subbands}
To obtain TGREED values, we employed a bank of temporal filters as described by (\ref{eqn:filter_bank}). In our implementation we used a 3-level wavelet decompositions resulting in 7 filters (a low pass filter is not used). In this experiment, we investigated the contribution of each subband by training and testing them individually. The results are shown in Fig. \ref{fig:subband_overall} where SROCC is plotted against the center frequencies (normalized to $[0,\pi]$ range) for each subband. From the plots we can infer that the middle frequency regions tended to have higher individual correlation as compared to other parts of the frequency spectrum. This behavior can be explained in terms of the temporal contrast sensitivity function (CSF) \cite{robson1966spatial} of human vision, according to which sensitivity to the visual signal is band-pass, resulting in reduced sensitivity to lower and higher frequencies. Note that the correlation values plotted along the y-axis in Fig. \ref{fig:subband_overall} are sensitive to the presence of temporal frequencies present in the dataset.


\begin{table*}
\caption{Comparison of compute times of various VQA models on 100 frames of $1920 \times 1080$ resolution video.}
\label{table:time_complexity}
\footnotesize
    \begin{tabular}{|c|c|c|c|c|c|c|c|c|c|c|c|c|}
        \hline
        \multirow{2}{*} & \multirow{2}{*}{PSNR} & \multirow{2}{*}{SSIM} & \multirow{2}{*}{MS-SSIM} & \multirow{2}{*}{FSIM} & \multirow{2}{*}{ST-RRED} & \multirow{2}{*}{SpEED} & \multirow{2}{*}{FRQM} & \multirow{2}{*}{deepVQA} & \multirow{2}{*}{VMAF} & GREED- & GREED- & GREED- \\ 
        ~ & ~ & ~ & ~ & ~ & ~ & ~ & ~ & ~ & ~ & Haar & db2 & bior2.2 \\ \hline
        Time (sec.) & 1.51 & 17.94 & 23.79 & 29.60 & 777.85 & 15.40 & 6.71 & 44.94 & 35.60 & 7.75 & 7.83 &  8.17 \\
        \hline
    \end{tabular}
\end{table*}


\begin{table}[t]
\caption{Performance comparison of different FR algorithms on the BVI-HFR Database.}
\label{table:BVI_HFR}
    \centering
    \begin{tabular}{|c||c|c|c|c|c|}
        \hline
        ~ & SROCC $\uparrow$ & KROCC $\uparrow$ & PLCC $\uparrow$ & RMSE $\downarrow$ \\ \hline \hline
        PSNR & 0.2552 & 0.1818 & 0.3155 & 17.04 \\ 
        SSIM \cite{wang2004image} & 0.1958 & 0.1515 & 0.3532 & 16.96 \\ 
        MS-SSIM \cite{wang2003multiscale} & 0.2063 & 0.1515 & 0.3583 & 16.96 \\ 
        FSIM \cite{zhang2011fsim} & 0.1888 & 0.1212 & 0.3448 & 17.16 \\ 
        ST-RRED \cite{soundararajan2012video} & 0.2028 & 0.1515 & 0.1699 & 17.97 \\
        SpEED \cite{bampis2017speed} & 0.2657 & 0.1818 & 0.2304 & 18.30 \\ 
        FRQM \cite{zhang2017frame} & \textbf{0.9021} & \textbf{0.7576} & \textbf{0.9394} & \textbf{6.37} \\ 
        VMAF \cite{VMAF2016} & 0.1888 & 0.1212 & 0.3703 & 16.59 \\
        deepVQA \cite{kim2018deep} & 0.1469 & 0.1212 & 0.2013 & 17.44 \\ \hline
        GREED-Haar & 0.7762 & 0.5455 & 0.8225 & 10.61 \\
        GREED-db2 & 0.7133 & 0.5152 & 0.772 & 11.54 \\
        GREED-bior2.2 & \textbf{0.8042} & \textbf{0.6061} & \textbf{0.8312} & \textbf{10.36} \\
        \hline
    \end{tabular}
\end{table}

\subsection{Performance Analysis on BVI-HFR Database}
BVI-HFR \cite{mackin2018study} is another HFR-VQA database available in the public domain, consisting of 22 source sequences spanning 4 frame rates: 15, 30, 60 and 120 fps. BVI-HFR and LIVE-YT-HFR differ in the downsampling scheme employed to obtain lower frame rate videos: the creators of BVI-HFR used temporal frame averaging, while in LIVE-YT-HFR, frame dropping was used. The choice of downsampling method can have a significant impact on the quality of lower frame rate videos: temporal frame averaging often introduces motion blur, while frame dropping may result in judder/strobing artifacts. In this experiment we investigated the performance of GREED on BVI-HFR, specifically its sensitivity to detect motion blur artifacts. Note that the BVI-HFR Database is primarily focused towards frame rate distortions like motion blur, and does not include other impairments such as compression, white noise etc. We again randomly split the dataset into 70\%, 15\%, and 15\% subsets for training, validation and testing respectively, and while ensuring no overlap between the contents across these sets. The above procedure was repeated 200 times and the median SROCC performance computed and compared in Table \ref{table:BVI_HFR}. Here, the FRQM index achieved the highest correlation against subjective judgments, while GREED-bior2.2 was second best among the compared FR-VQA models. This suggests that GREED is not very sensitive to the choice of downsampling scheme used when obtaining lower frame rate sequences.

\begin{table}[t]
\caption{SROCC performance comparison on multiple VQA databases. The reported numbers are median values from every possible combination of train-test splits with 80\% of content used for training.}
    \label{Table:VQA_database}
    \centering
    \footnotesize
    \begin{tabular}{|c||c|c|c|}
        \hline
        ~    & LIVE-VQA  & LIVE-mobile & CSIQ-VQA \\ \hline \hline 
        PSNR & 0.711 & 0.788 & 0.601 \\ 
        SSIM \cite{wang2004image} & 0.802 & 0.798 & 0.705 \\ 
        MS-SSIM \cite{wang2003multiscale} & \textbf{0.830} & 0.800 & 0.757 \\ 
        FSIM \cite{zhang2011fsim} & 0.806 & 0.868 & 0.752 \\ 
        ST-RRED \cite{soundararajan2012video} & \textbf{0.826} & 0.882 & \textbf{0.813} \\ 
        SpEED \cite{bampis2017speed} & 0.801 & \textbf{0.886} & 0.743 \\ 
        VMAF \cite{VMAF2016}& 0.794 & \textbf{0.897} & 0.618 \\
        deepVQA \cite{kim2018deep} & - & 0.826 & 0.702 \\ \hline
        GREED-Haar & 0.771 & 0.875 & 0.785 \\
        GREED-db2 & 0.735 & 0.850 & \textbf{0.786} \\
        GREED-bior2.2 & 0.750 & 0.863 & 0.780 \\
        \hline
    \end{tabular}
\end{table}

\subsection{Performance Comparison on other VQA databases}
We further investigated the generalizability of the GREED features by evaluating them on three popular VQA databases: LIVE-VQA \cite{seshadrinathan2010study}, LIVE-mobile \cite{moorthy2012video} and CSIQ-VQA \cite{vu2014vis3}. These databases contain videos of the same frame rate for both reference and distorted sequences, thus the TGREED term in (\ref{eqn:GTI}) will only depend on the absolute difference term, since the ratio term reduces to unity. On each database, we divided the contents into non-overlapping 80\% and 20\% subsets for training and testing, respectively. Further, this procedure was repeated for all possible train-test combinations, and median SROCC performance reported in Table \ref{Table:VQA_database}. From the Table, we observe that GREED models achieved comparable performance to state-of-the-art (SOTA) VQA methods. This also indicates the efficacy of the features employed by GREED is not restricted to HFR content, and generalizes well across other types of artifacts which may arise in non-HFR streaming and social media scenarios. 

To analyze the dependence of GREED on training data, we also performed cross dataset evaluation, whereby we trained on one database and used the remaining datasets for testing. The results of this experiment are given in Table \ref{table:cross_data} where the correlation remains nearly unchanged regardless of the training data employed, highlighting the robustness of the features used in GREED. 

\begin{table}[t!]
\caption{Cross database SROCC performance comparison of GREED. The highest value in each column is boldfaced.}
    \label{table:cross_data}
\footnotesize
\center
    \subfloat[GREED-Haar]{
    \begin{tabular}{|c||c|c|c|c|} 
    \hline
    \multirow{2}{*}{\backslashbox{Train}{Test}} & LIVE- & LIVE  & LIVE & CSIQ \\
    ~ & YT-HFR & VQA  & mobile & VQA \\\hline \hline
    LIVE-YT-HFR & - & \textbf{0.705} & \textbf{0.804} & 0.626 \\
    LIVE-VQA & 0.705 & - & 0.801 & 0.622\\
    LIVE-mobile & 0.658 & 0.692 & - & \textbf{0.636}\\
    CSIQ-VQA & \textbf{0.719} & 0.682 & 0.797 & -\\
    \hline
    \end{tabular}} \\
    
    \subfloat[GREED-db2]{   
    \begin{tabular}{ |c||c|c|c|c| } 
    \hline
    \multirow{2}{*}{\backslashbox{Train}{Test}} & LIVE- & LIVE  & LIVE & CSIQ \\
    ~ & YT-HFR & VQA  & mobile & VQA \\\hline \hline
    LIVE-YT-HFR & - & \textbf{0.685} & 0.807 & \textbf{0.634} \\
    LIVE-VQA & 0.705 & - & 0.807 & 0.621\\
    LIVE-mobile & 0.660 & 0.664 & - & 0.617\\
    CSIQ-VQA & \textbf{0.721} & 0.678 & \textbf{0.820} & -\\
    \hline
    \end{tabular}} \\
    
    \subfloat[GREED-bior2.2]{
    \begin{tabular}{ |c||c|c|c|c| } 
    \hline
    \multirow{2}{*}{\backslashbox{Train}{Test}} & LIVE- & LIVE  & LIVE & CSIQ \\
    ~ & YT-HFR & VQA  & mobile & VQA \\\hline \hline
    LIVE-YT-HFR & - & 0.697 & 0.798 & \textbf{0.616} \\
    LIVE-VQA & 0.707 & - & 0.825 & 0.600 \\
    LIVE-mobile & 0.678 & 0.599 & - & 0.418\\
    CSIQ-VQA & \textbf{0.738} & \textbf{0.716} & \textbf{0.837} & -\\
    \hline
    \end{tabular}} 
\end{table}

\begin{table*}[t]
	\caption{SROCC performance comparison of HFR-VMAF for individual frame rates on the LIVE-YT-HFR Database. In each column the best value is marked in boldface.}
	\label{Table:HFR_VMAF}
	\centering
	\footnotesize
	\scalebox{1}{
		\begin{tabular}{|c||c|c|c|c|c|c|c|}
			\hline
			~ & 24 fps & 30 fps & 60 fps & 82 fps & 98 fps &  120 fps & Overall \\ \hline \hline 
			VMAF \cite{VMAF2016} & 0.2500 & 0.3625 & 0.6304 & 0.7339 & 0.8607 & 0.8182 & 0.7782 \\ 
			GREED-Haar & 0.6196 & 0.5482 & 0.7125 & 0.7464 & 0.8054 & 0.8112 & 0.8305 \\
			GREED-db2 & 0.6696 & 0.6179 & 0.6982 & 0.7250 & 0.7518 & 0.8322 & 0.8347 \\
			GREED-bior2.2 & 0.7268 & 0.7018 & 0.7321 & 0.8179 & 0.8643 & \textbf{0.8881} & \textbf{0.8822}\\
			\hline
			HFR-VMAF-Haar & 0.6946 & 0.6357 & 0.7857 & 0.8161 & 0.8571 & 0.8392 & 0.7608 \\
			HFR-VMAF-db2 & 0.7107 & 0.6804 & 0.7929 & 0.8286 & 0.8536 & 0.8601 & 0.7773 \\
			HFR-VMAF-bior2.2 & \textbf{0.7714} & \textbf{0.7429} & \textbf{0.7893} & \textbf{0.8536} & \textbf{0.8821} & \textbf{0.8881} & 0.8160\\
			\hline
		\end{tabular}
	}
\end{table*}

\subsection{Time Complexity}
In Table \ref{table:time_complexity} we compare the compute times (in seconds) of various VQA models on 100 frames of video having $1920 \times 1080$ resolution. The compute times were calculated on a Xeon E5 2620 v4 2.1 GHz CPU with 64 GB RAM. As compared to other VQA models, GREED has relatively low complexity, as indicated by the lower time required for feature computation. The computational complexity of GREED is mainly dependent on the evaluation of TGREED features, since TGREED contributes 14 of the 16 features employed in GREED. 

\section{HFR-VMAF}
VMAF video quality prediction framework has demonstrated high prediction performance when the reference and distorted videos have the same frame rate. Because of this, VMAF has been widely employed by Netflix to control the quality of its streaming content. Given that VMAF achieves competitive performance when the videos being compared have the same frame rates, we attempted to leverage this usefulness by combining GREED and VMAF predictions. Specifically, we introduce a variant of VMAF that we dub as HFR-VMAF, defined as the average of VMAF (here we use $100 - \text{VMAF}$ since we require the score of pristine video to be at zero) and GREED predictions:
\begin{align}
    \text{HFR-VMAF}(R,D) = \frac{1}{2}(\text{VMAF}(PR,D) + \text{GREED}(R,D))
    \label{eqn:HFR_VMAF}
\end{align}
Note that VMAF is computed between the distorted video $D$ and subsampled reference $PR$ sequences, unlike the temporally upsampled distorted video $D$ used in Tables \ref{Table:MOS_comparison} and \ref{Table:FPS_comparison}. Although subsampling the reference video can result in temporal artifacts like judder, strobing etc., we empirically observed that when combined with GREED, these artifacts tended to have negligible effect on the quality predictions. Moreover, we observed that the quality predictions produced by HFR-VMAF to be highly effective, particularly when sets of videos having fixed frame rates were compared. This is illustrated in Table \ref{Table:HFR_VMAF}, where we observed considerable improvement in correlation values on individual frame rates. Yet, although there is a performance boost when fixed frame rates were considered, there was also considerable correlation degradation when entire database was included, indicating that HFR-VMAF is less efficient at differentiating the perceptual quality of videos over multiple frame rates. This suggests that HFR-VMAF could be beneficial when videos of the same frame rates are compared, while GREED is more suitable when videos across different frame rates are compared. 

\section{Conclusion and Future Work}
\label{sec:conclusion}
We proposed a new model that accurately predicts frame rate dependent video quality based on measurement of bandpass video statistics. A distinguishing element of the new GREED model is that it can be used to measure video quality when reference and distorted sequences have differing frame rates, with no requirement of any temporal pre-processing. We conducted a comprehensive and holistic evaluation of GREED against human judgments of video quality and found that GREED delivers more accurate and robust predictions of quality than other VQA models on variable frame rate videos. We conducted ablation studies to analyze the significance of the spatial and temporal components of GREED, and demonstrated their complementary nature in capturing relevant perceptual information. We evaluated the generalizability of GREED features on multiple fixed frame rate VQA databases and observed comparable performance to SOTA VQA models. We also proposed HFR-VMAF, an extension of VMAF to HFR videos incorporating the advantages of both the GREED and VMAF methods. HFR-VMAF was observed to enhance prediction performance when videos of fixed frame rates were analyzed. A software release of GREED has been made available online\footnote{\url{https://github.com/pavancm/GREED}}.

Although GREED achieves high correlations against perceptual judgments, we observed a shortcoming that could affect prediction performance. The differences between the frame rates of reference and distorted videos can influence performance, as observed in Table \ref{Table:FPS_comparison} where lower frame rate videos led to worse performance than higher frame rate videos. Although HFR-VMAF addresses this concern, it results in performance degradations when videos across different frame rates are considered. 
A more careful design addressing the above drawback would be beneficial towards understanding frame rate influences on HFR video quality, and in creating further improved models.

\ifCLASSOPTIONcaptionsoff
  \newpage
\fi

\bibliographystyle{IEEEtran}
\bibliography{template_4}

\begin{thebibliography}{10}
\providecommand{\url}[1]{#1}
\csname url@samestyle\endcsname
\providecommand{\newblock}{\relax}
\providecommand{\bibinfo}[2]{#2}
\providecommand{\BIBentrySTDinterwordspacing}{\spaceskip=0pt\relax}
\providecommand{\BIBentryALTinterwordstretchfactor}{4}
\providecommand{\BIBentryALTinterwordspacing}{\spaceskip=\fontdimen2\font plus
\BIBentryALTinterwordstretchfactor\fontdimen3\font minus
  \fontdimen4\font\relax}
\providecommand{\BIBforeignlanguage}[2]{{%
\expandafter\ifx\csname l@#1\endcsname\relax
\typeout{** WARNING: IEEEtran.bst: No hyphenation pattern has been}%
\typeout{** loaded for the language `#1'. Using the pattern for}%
\typeout{** the default language instead.}%
\else
\language=\csname l@#1\endcsname
\fi
#2}}
\providecommand{\BIBdecl}{\relax}
\BIBdecl

\bibitem{nasiri2015perceptual}
R.~M. Nasiri, J.~Wang, A.~Rehman, S.~Wang, and Z.~Wang, ``Perceptual quality
  assessment of high frame rate video,'' in \emph{IEEE International Workshop
  on Multimedia Signal Processing (MMSP)}.\hskip 1em plus 0.5em minus
  0.4em\relax IEEE, 2015, pp. 1--6.

\bibitem{mackin2018study}
A.~Mackin, F.~Zhang, and D.~R. Bull, ``A study of high frame rate video
  formats,'' \emph{IEEE Trans. Multimedia}, vol.~21, no.~6, pp. 1499--1512,
  2018.

\bibitem{pavan2020liveythfr}
P.~C. Madhusudana, X.~Yu, N.~Birkbeck, Y.~Wang, B.~Adsumilli, and A.~C. Bovik,
  ``Subjective and objective quality assessment of high frame rate videos,''
  \emph{IEEE Access}, vol.~9, 2021.

\bibitem{soundararajan2012video}
R.~{Soundararajan} and A.~C. {Bovik}, ``Video quality assessment by reduced
  reference spatio-temporal entropic differencing,'' \emph{IEEE Trans. Circuits
  Syst. Video Technol.}, vol.~23, no.~4, pp. 684--694, April 2013.

\bibitem{VMAF2016}
Z.~Li, A.~Aaron, I.~Katsavounidis, A.~Moorthy, and M.~Manohara, ``{Toward a
  practical perceptual video quality metric},''
  \url{http://techblog.netflix.com/2016/06/toward-practical-perceptual-video.html}.

\bibitem{bampis2017speed}
C.~G. {Bampis}, P.~{Gupta}, R.~{Soundararajan}, and A.~C. {Bovik},
  ``{SpEED-QA}: Spatial efficient entropic differencing for image and video
  quality,'' \emph{IEEE Signal Process. Lett.}, vol.~24, no.~9, pp. 1333--1337,
  Sep. 2017.

\bibitem{bampis2018spatiotemporal}
C.~G. Bampis, Z.~Li, and A.~C. Bovik, ``Spatiotemporal feature integration and
  model fusion for full reference video quality assessment,'' \emph{IEEE Trans.
  Circuits Syst. Video Technol.}, vol.~29, no.~8, pp. 2256--2270, 2018.

\bibitem{chikkerur2011objective}
S.~Chikkerur, V.~Sundaram, M.~Reisslein, and L.~J. Karam, ``Objective video
  quality assessment methods: A classification, review, and performance
  comparison,'' \emph{IEEE Trans. Broadcast.}, vol.~57, no.~2, pp. 165--182,
  2011.

\bibitem{wang2004image}
Z.~{Wang}, A.~C. {Bovik}, H.~R. {Sheikh}, and E.~P. {Simoncelli}, ``Image
  quality assessment: from error visibility to structural similarity,''
  \emph{IEEE Trans. Image Process.}, vol.~13, no.~4, pp. 600--612, April 2004.

\bibitem{wang2003multiscale}
Z.~{Wang}, E.~P. {Simoncelli}, and A.~C. {Bovik}, ``Multiscale structural
  similarity for image quality assessment,'' in \emph{Asilomar Conf. Signals
  Syst. Comput.}, vol.~2, Nov 2003, pp. 1398--1402 Vol.2.

\bibitem{zhang2011fsim}
L.~Zhang, L.~Zhang, X.~Mou, and D.~Zhang, ``{FSIM}: A feature similarity index
  for image quality assessment,'' \emph{IEEE Trans. Image Process.}, vol.~20,
  no.~8, pp. 2378--2386, 2011.

\bibitem{pinson2004new}
M.~H. {Pinson} and S.~{Wolf}, ``A new standardized method for objectively
  measuring video quality,'' \emph{IEEE Trans. Broadcast.}, vol.~50, no.~3, pp.
  312--322, Sep. 2004.

\bibitem{pinson2014temporal}
M.~H. Pinson, L.~K. Choi, and A.~C. Bovik, ``Temporal video quality model
  accounting for variable frame delay distortions,'' \emph{IEEE Trans.
  Broadcast.}, vol.~60, no.~4, pp. 637--649, 2014.

\bibitem{moorthy2012video}
A.~K. Moorthy, L.~K. Choi, A.~C. Bovik, and G.~De~Veciana, ``Video quality
  assessment on mobile devices: Subjective, behavioral and objective studies,''
  \emph{IEEE J. Sel. Topics Signal Process.}, vol.~6, no.~6, pp. 652--671,
  2012.

\bibitem{seshadrinathan2009motion}
K.~{Seshadrinathan} and A.~C. {Bovik}, ``Motion tuned spatio-temporal quality
  assessment of natural videos,'' \emph{IEEE Trans. Image Process.}, vol.~19,
  no.~2, pp. 335--350, Feb 2010.

\bibitem{seshadrinathan2007structural}
K.~Seshadrinathan and A.~C. Bovik, ``A structural similarity metric for video
  based on motion models,'' in \emph{IEEE Intl. Conf. Acoustics, Speech, and
  Signal Processing}, 2007, pp. I--869.

\bibitem{vu2011spatiotemporal}
P.~V. {Vu}, C.~T. {Vu}, and D.~M. {Chandler}, ``A spatiotemporal
  most-apparent-distortion model for video quality assessment,'' in \emph{IEEE
  Int’l Conf. Image Process.}, Sep. 2011, pp. 2505--2508.

\bibitem{larson2010most}
E.~C. Larson and D.~M. Chandler, ``Most apparent distortion: full-reference
  image quality assessment and the role of strategy,'' \emph{Journal of
  Electronic Imaging}, vol.~19, no.~1, p. 011006, 2010.

\bibitem{you2013attention}
J.~{You}, T.~{Ebrahimi}, and A.~{Perkis}, ``Attention driven foveated video
  quality assessment,'' \emph{IEEE Trans. Image Process.}, vol.~23, no.~1, pp.
  200--213, Jan 2014.

\bibitem{ortiz2014full}
B.~{Ortiz-Jaramillo}, A.~{Kumcu}, L.~{Platisa}, and W.~{Philips}, ``A full
  reference video quality measure based on motion differences and saliency maps
  evaluation,'' in \emph{Int'l Conf. on Comp. Vision Theory and Applications},
  vol.~2, Jan 2014, pp. 714--722.

\bibitem{manasa2016optical}
K.~{Manasa} and S.~S. {Channappayya}, ``An optical flow-based full reference
  video quality assessment algorithm,'' \emph{IEEE Trans. Image Process.},
  vol.~25, no.~6, pp. 2480--2492, June 2016.

\bibitem{kim2018deep}
W.~Kim, J.~Kim, S.~Ahn, J.~Kim, and S.~Lee, ``Deep video quality assessor: From
  spatio-temporal visual sensitivity to a convolutional neural aggregation
  network,'' in \emph{Proc. European Conf. Comput. Vision}, September 2018, pp.
  219--234.

\bibitem{becker2019neural}
S.~Becker, K.-R. M{\"u}ller, T.~Wiegand, and S.~Bosse, ``A neural network model
  of spatial distortion sensitivity for video quality estimation,'' in
  \emph{IEEE Int'l Workshop on Machine Learning for Signal Processing}, 2019,
  pp. 1--6.

\bibitem{xu2020c3dvqa}
M.~Xu, J.~Chen, H.~Wang, S.~Liu, G.~Li, and Z.~Bai, ``{C3DVQA}: Full-reference
  video quality assessment with 3d convolutional neural network,'' in
  \emph{IEEE Int'l Conf. on Acoustics, Speech and Signal Process.}, 2020, pp.
  4447--4451.

\bibitem{nasiri2017perceptual}
R.~M. Nasiri and Z.~Wang, ``Perceptual aliasing factors and the impact of frame
  rate on video quality,'' in \emph{IEEE Int’l Conf. Image Process.}, 2017,
  pp. 3475--3479.

\bibitem{nasiri2018temporal}
R.~M. Nasiri, Z.~Duanmu, and Z.~Wang, ``Temporal motion smoothness and the
  impact of frame rate variation on video quality,'' in \emph{IEEE Int’l
  Conf. Image Process.}, 2018, pp. 1418--1422.

\bibitem{zhang2017frame}
F.~{Zhang}, A.~{Mackin}, and D.~R. {Bull}, ``A frame rate dependent video
  quality metric based on temporal wavelet decomposition and spatiotemporal
  pooling,'' in \emph{IEEE Int’l Conf. Image Process.}, Sep. 2017, pp.
  300--304.

\bibitem{pavan2020gsti}
P.~C. {Madhusudana}, N.~{Birkbeck}, Y.~{Wang}, B.~{Adsumilli}, and A.~C.
  {Bovik}, ``Capturing video frame rate variations via entropic differencing,''
  \emph{IEEE Sig. Process. Letters}, vol.~27, pp. 1809--1813, 2020.

\bibitem{saad2014blind}
M.~A. Saad, A.~C. Bovik, and C.~Charrier, ``Blind prediction of natural video
  quality,'' \emph{IEEE Trans. Image Process.}, vol.~23, no.~3, pp. 1352--1365,
  March 2014.

\bibitem{sheikh2006image}
H.~R. Sheikh and A.~C. Bovik, ``Image information and visual quality,''
  \emph{IEEE Trans. Image Process.}, vol.~15, no.~2, pp. 430--444, Feb. 2006.

\bibitem{soundararajan2012rred}
R.~Soundararajan and A.~C. Bovik, ``{RRED} indices: Reduced reference entropic
  differencing for image quality assessment,'' \emph{IEEE Trans. Image
  Process.}, vol.~21, no.~2, pp. 517--526, Feb. 2012.

\bibitem{ruderman1994statistics}
D.~L. Ruderman, ``The statistics of natural images,'' \emph{Network:
  computation in neural systems}, vol.~5, no.~4, pp. 517--548, 1994.

\bibitem{mittal2012no}
A.~Mittal, A.~K. Moorthy, and A.~C. Bovik, ``No-reference image quality
  assessment in the spatial domain,'' \emph{IEEE Trans. Image Process.},
  vol.~21, no.~12, pp. 4695--4708, Dec. 2012.

\bibitem{mittal2013making}
A.~Mittal, R.~Soundararajan, and A.~C. Bovik, ``Making a ``completely blind''
  image quality analyzer,'' \emph{IEEE Sig. Process. Letters}, vol.~20, no.~3,
  pp. 209--212, Mar. 2013.

\bibitem{chang2000adaptive}
S.~G. Chang, B.~Yu, and M.~Vetterli, ``Adaptive wavelet thresholding for image
  denoising and compression,'' \emph{IEEE Trans. Image Process.}, vol.~9,
  no.~9, pp. 1532--1546, 2000.

\bibitem{do2002wavelet}
M.~N. Do and M.~Vetterli, ``Wavelet-based texture retrieval using generalized
  gaussian density and {K}ullback-{L}eibler distance,'' \emph{IEEE Trans. Image
  Process.}, vol.~11, no.~2, pp. 146--158, 2002.

\bibitem{zhao2004sum}
Q.~Zhao, H.-w. Li, and Y.-t. Shen, ``On the sum of generalized gaussian random
  signals,'' in \emph{IEEE Int'l Conf. on Signal Process.}, 2004, pp. 50--53.

\bibitem{soury2015new}
H.~Soury and M.-S. Alouini, ``New results on the sum of two generalized
  {G}aussian random variables,'' in \emph{IEEE Global Conf. on Signal and
  Information Process.}, 2015, pp. 1017--1021.

\bibitem{pan2012exposing}
X.~Pan, X.~Zhang, and S.~Lyu, ``Exposing image splicing with inconsistent local
  noise variances,'' in \emph{IEEE International Conference on Computational
  Photography (ICCP)}, 2012, pp. 1--10.

\bibitem{ffmpeg}
{FFmpeg}, ``{Encoding for streaming sites},''
  \url{https://trac.ffmpeg.org/wiki}, [Online; accessed 1-November-2019].

\bibitem{scholkopf2000new}
B.~Sch{\"o}lkopf, A.~J. Smola, R.~C. Williamson, and P.~L. Bartlett, ``New
  support vector algorithms,'' \emph{Neural {C}omput.}, vol.~12, no.~5, pp.
  1207--1245, 2000.

\bibitem{chang2011libsvm}
C.-C. Chang and C.-J. Lin, ``{LIBSVM}: A library for support vector machines,''
  \emph{ACM Trans. Intell. Syst. Technol.}, vol.~2, no.~3, pp. 27:1--27:27, May
  2011.

\bibitem{coifman1992entropy}
R.~R. Coifman and M.~V. Wickerhauser, ``Entropy-based algorithms for best basis
  selection,'' \emph{IEEE Trans. Inf. Theory}, vol.~38, no.~2, pp. 713--718,
  1992.

\bibitem{seshadrinathan2010study}
K.~Seshadrinathan, R.~Soundararajan, A.~C. Bovik, and L.~K. Cormack, ``Study of
  subjective and objective quality assessment of video,'' \emph{IEEE Trans.
  Image Process.}, vol.~19, no.~6, pp. 1427--1441, June 2010.

\bibitem{VQEG2000}
VQEG, ``Final report from the video quality experts group on the validation of
  objective quality metrics for video quality assessment,'' 2000.

\bibitem{robson1966spatial}
J.~G. Robson, ``Spatial and temporal contrast-sensitivity functions of the
  visual system,'' \emph{J. Opt. Soc. Amer.}, vol.~56, no.~8, pp. 1141--1142,
  1966.

\bibitem{vu2014vis3}
P.~V. Vu and D.~M. Chandler, ``Vis3: an algorithm for video quality assessment
  via analysis of spatial and spatiotemporal slices,'' \emph{Journal of
  Electronic Imaging}, vol.~23, no.~1, p. 013016, 2014.

\end{thebibliography}

\begin{IEEEbiography}[{\includegraphics[width=1in,height=1.25in,clip,keepaspectratio]{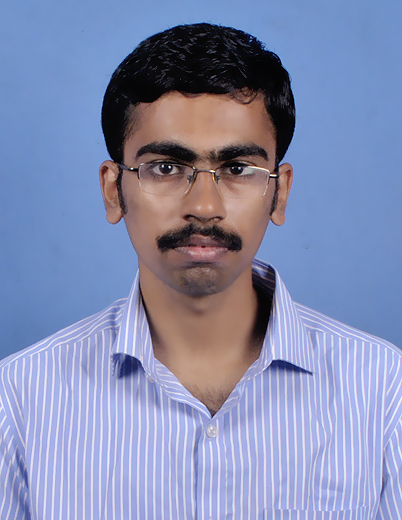}}]{Pavan C. Madhusudana}
received the B.Tech. degree in Electronics and Communication Engineering from The National Institute of Technology Karnataka (NITK), Surathkal, India, in 2016, and the M.Tech. (Research) degree in Electrical and Communication Engineering from the Indian Institute of Science (IISc), Bangalore, India in 2018. He is currently pursuing the Ph.D. degree in Electrical and Computer engineering with The University of Texas at Austin, USA. His research interests include image and video signal processing, computer vision, and machine learning.
\end{IEEEbiography}

\begin{IEEEbiography}
[{\includegraphics[width=1in,height=1.25in,clip,keepaspectratio]{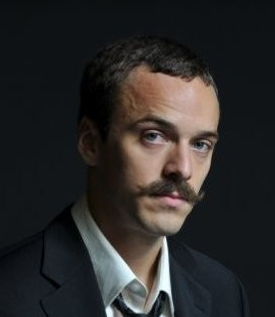}}]{Neil Birkbeck} obtained his Ph.D from the University of Alberta in 2011 working on topics in computer vision, graphics and robotics, with a specific focus on image-based modeling and rendering. He went on to become a Research Scientist at Siemens corporate research working on automatic detection and segmentation of anatomical structures in full body medical images. He is now a software engineer in the Media Algorithms team at YouTube/Google, with research interests in perceptual video processing, video coding, and video quality assessment.
\end{IEEEbiography}

\begin{IEEEbiography}
[{\includegraphics[width=1in,height=1.25in,clip,keepaspectratio]{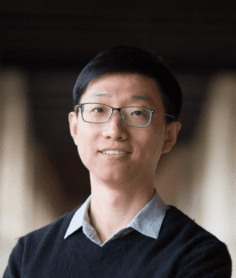}}]{Yilin Wang}
received B.S. and M.S. degrees in Computer Science from Nanjing University, China, in 2005 and 2008 respectively, PhD degree in Computer Science from the University of North Carolina at Chapel Hill in 2014, working on topics in computer vision and image processing. After graduation, he joined the Media Algorithm team in Youtube/Google. His research fields include video processing infrastructure, video quality assessment, and video compression.
\end{IEEEbiography}

\begin{IEEEbiography}
[{\includegraphics[width=1in,height=1.25in,clip,keepaspectratio]{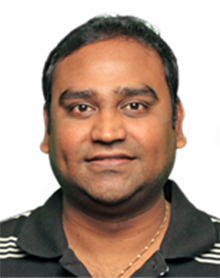}}]{Balu Adsumilli} manages and leads the Media Algorithms group at YouTube/Google. He did his masters in University of Wisconsin Madison in 2002, and his PhD at University of California Santa Barbara in 2005, on watermark-based error resilience in video communications. From 2005 to 2011, he was Sr. Research Scientist at Citrix Online, and from 2011-2016, he was Sr. Manager Advanced Technology at GoPro, at both places developing algorithms for images/video quality enhancement, compression, capture, and streaming. He is an active member of IEEE (and MMSP TC), ACM, SPIE, and VES, and has co-authored more than 120 papers and patents. His fields of research include image/video processing, machine vision, video compression, spherical capture, VR/AR, visual effects, and related areas.
\end{IEEEbiography}

\begin{IEEEbiography}[{\includegraphics[width=1in,height=1.25in,clip,keepaspectratio]{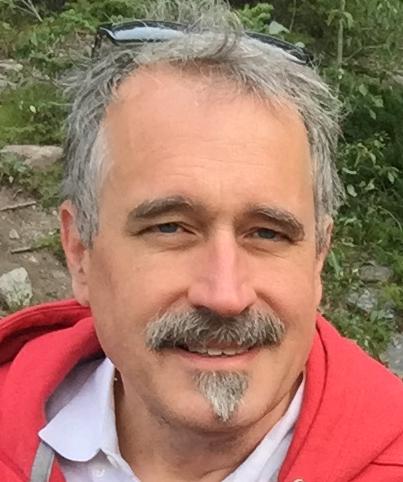}}]{Alan C. Bovik} (F ’95) is the Cockrell Family Regents Endowed Chair Professor at The University of Texas at Austin. His research interests include image processing, digital photography, digital television, digital streaming video, social media, and visual perception. For his work in these areas he has been the recipient of the 2019 Progress Medal from The Royal Photographic Society, the 2019 IEEE Fourier Award, the 2017 Edwin H. Land Medal from The Optical Society, a 2015 Primetime Emmy Award for Outstanding Achievement in Engineering Development from the Television Academy, a 2020 Technology and Engineering Emmy Award from the National Academy for Television Arts and Sciences, and the Norbert Wiener Society Award and the Karl Friedrich Gauss Education Award from the IEEE Signal Processing Society. He has also received about 10 ‘best journal paper’ awards, including the 2016 IEEE Signal Processing Society Sustained Impact Award. His books include The Essential Guides to Image and Video Processing. He co-founded and was longest-serving Editor-in-Chief of the IEEE Transactions on Image Processing, and also created/Chaired the IEEE International Conference on Image Processing which was first held in Austin, Texas, 1994.
\end{IEEEbiography}

\end{document}